\documentclass[11pt]{article}
\usepackage{amsmath}
\usepackage{xspace}
\usepackage{amssymb}
\usepackage{epsfig}

\newtheorem{Thm}{Theorem}
\newtheorem{Lem}[Thm]{Lemma}

\newenvironment{proof}{\noindent {\textbf{Proof }}}{$\Box$ \medskip}

\newcommand\np{\mbox{$\mathbf{NP}$}\xspace}
\newcommand\bpp{\mbox{$\mathbf{BPP}$}\xspace}
\newcommand\bqp{\mbox{$\mathbf{BQP}$}\xspace}
\newcommand\qma{\mbox{$\mathbf{QMA}$}\xspace}

\newcommand\B{\{0,1\}}
\newcommand\ket[1]{| #1 \rangle}
\newcommand\bra[1]{\langle #1 |}
\newcommand\qip[2]{\langle #1 | #2 \rangle}
\newcommand\pr{\mbox{\bf Pr}}
\newcommand\av{\mbox{\bf{\bf E}}}
\newcommand\var{\mbox{\bf Var}}
\newcommand{\ceil}[1]{\lceil #1 \rceil}

\newcommand {\ie} {\emph{i.e.}\xspace}
\newcommand {\st} {\emph{s.t.}\xspace}

\begin{document}
\title{\textbf{Several natural BQP-Complete problems}\thanks{P.W. is
supported in part by the Army Research Office under Grant
No.~W911NF-05-1-0294 and by the National Science Foundation under
Grant No.~PHY-456720.  S.Z. is supported in part by NSF grants
CCR-0310466 and CCF-0426582.}}  \author{Pawel Wocjan
\thanks{Computer Science Department \& Institute for Quantum
Information, California Institute of Technology, Pasadena, CA 91125,
USA. Email: wocjan@cs.caltech.edu} \quad and \quad Shengyu Zhang
\thanks{Computer Science Department, Princeton University. 35 Olden St, Princeton NJ
08540, USA.  Email: szhang@cs.princeton.edu}} \date{}

\maketitle

\abstract{A central problem in quantum computing is to identify
computational tasks which can be solved substantially faster on a
quantum computer than on any classical computer.  By studying the
hardest such tasks, known as BQP-complete problems, we deepen our
understanding of the power and limitations of quantum computers. We
present several \bqp-complete problems, including Phase Estimation
Sampling and Local Hamiltonian Eigenvalue Sampling. Different than the
previous known \bqp complete problems (the Quadratically Signed Weight
Enumerator problem \cite{KL01} and the Approximation of Jones
Polynomials \cite{FKW02, FLW02, AJL06}), our problems are of a basic
linear algebra nature and are closely related to the well-known
quantum algorithm and quantum complexity theories.}

\section{Introduction}
The last decade has witnessed a great surge of fruitful studies in
the new paradigm of quantum computing, especially after Shor
discovered the celebrated quantum algorithms for factoring and
discrete logarithm \cite{Sh97}. While many other areas like quantum
complexity theory, quantum cryptography and quantum error correction
have been rapidly developed, the progress in quantum algorithm
design, one of the most central tasks of quantum computing, appears
much slower than what people had expected. Especially, designing
quantum algorithms that have super-polynomial speedup over their
classical counterparts seems very hard, and we have only succeeded
on few problems, including Hallgren's polynomial time algorithm for
Pell's equation and class groups \cite{Ha02}, van Dam, Hallgren and
Ip's polynomial time algorithm for some hidden shift problems
\cite{vDHI03}, Kuperberg's subexponential time algorithm for HSP
over the dihedral group \cite{Ku03}, and some others. In a recent
survey by Shor \cite{Sh04}, two major possible reasons are proposed
to explain the difficulty. One is that we have no good intuitions to
design quantum algorithms due to our lack of quantum experience. The
other is rather pessimistic: there may actually be only a few
problems for which quantum computers have significant advantages
over classical computers, and probably we have already found almost
all of them.

Considering the enormous payoff of efficient quantum algorithms and
these possible reasons for the difficulties to discover them, it is
central to \emph{identify} the class of problems that quantum
computers can have substantial speedup over classical
computers. Stated in the language of complexity theory, the task
amounts to pinning down \bqp, the class of languages decidable by a
uniform family of polynomial-size quantum circuits with bounded
error. While the class could be equivalently characterized by various
models, including the quantum Turing machine \cite{BV97}, quantum
circuits \cite{Ya93} and quantum adiabatic computation \cite{FGGS00,
AvDK+04}, identifying it is difficult, just as it is difficult to
characterize other important classes such as \np. Nevertheless, it was
recently found that the Quadratically Signed Weight Enumerator problem
\cite{KL01} and the problem of approximately evaluating the Jones
polynomial \cite{FKW02, FLW02, AJL06} are
\bqp-complete\footnote{Strictly speaking, approximation of the Jones
polynomial is neither a language nor a decision problem. Here the
concept of completeness is generalized in both their and our work: A
computational task is \bqp-complete if it can be solved by a
polynomial time quantum computer, and any \bqp problem can be reduced
to this problem.}.
Such problems are the hardest problems in \bqp. Consequently, the
study of \bqp-complete problems is important as it helps us to deepen
our understanding of \bqp by fully capturing the power of quantum
computation. This is similar to the study of \np-complete problem
which aims at understanding the power of nondeterminism.

In this paper, we provide several other \bqp-complete problems,
including the Local Hamiltonian Eigenvalue Sampling (LHES) problem
and the Phase Estimation Sampling (PES) problem, which are (roughly)
to sample an eigenvalue of a given Local Hamiltonian or a unitary
matrix. Both problems are of an elementary linear algebra nature,
which may make them accessible to more computer scientists. More
importantly, these two problems are very closely related to
well-known quantum algorithm and quantum complexity theories. First,
the Local Hamiltonian problem, which basically estimates the minimal
eigenvalue of the given local Hamiltonian, is well known to be
\qma-complete \cite{KSV02, KR03, KKR06}\footnote{Recently another
problem (Consistency of Local Density Matrices) is showed to be also
\qma-complete \cite{Liu06}.}. Our result on LHES says that if we do
not aim at estimating the minimal eigenvalue, but at sampling an
eigenvalue according to a very natural distribution, then the
problem becomes \bqp-complete. Second, Phase Estimation is a general
framework to design and study quantum algorithms. Many known quantum
algorithms that have exponential speedup over their classical
counterparts can be described in terms of the phase estimation
algorithm \cite{Ki95, CEMM98}. Now our result implies that this is
not an accident; actually all efficient quantum algorithms can be
done in this framework.

Measuring the energies of a quantum observable --- or in a
mathematical language, measuring eigenvalues of a Hamiltonian --- is
one of the most important tasks in quantum physics. The problems
related to estimating/approximating eigenvalues of Hamiltonians or
unitaries has been considered in the literature, including how to
make the measurement for special cases \cite{AL00, JB01, Ja02}, or
using the measurement to solve some computational tasks
\cite{WJDB04, CDF+02}. In this paper, we provide a precise
complexity-theoretic framework for studying these problems, and show
that the sampling versions of these two basic problems are actually
\bqp-complete.

The remainder of this paper is organized as follows. Section
\ref{sec: preliminaries} is devoted to basic notations used in the
paper, and the definitions of the three problems we are studying in
this paper. In Section \ref{sec: hardness}, we prove that the three
problems are \bqp-hard. In Section \ref{sec: in BQP}, we show that
the problems are in \bqp. The paper concludes with Section \ref{sec:
conclusion}, which mentions some open problems for future work.

\section{Preliminaries and definitions} \label{sec: preliminaries}
A quantum register consisting of $n$ qubits is described
mathematically by a tensor product Hilbert space ${\cal H} =
(\mathbb{C}^2)^{\otimes n}$.  A state of the quantum register is given
by a unit vector $\ket{\psi}$. Transformations of the states are
described by unitary operations acting on ${\cal H}$. Every unitary
operation has to be composed out of elementary gates for
implementation purposes. A measurement with respect to the orthogonal
subspaces $S_1$, $S_2$ (where $S_1 \oplus S_2 = \mathcal{H}$) causes
the system to collapse to $P_1 \ket{\psi}/\|P_1 \ket{\psi}\|$ or
$P_2\ket{\psi}/\|P_1 \ket{\psi}\|$, with probability $\|P_1
\ket{\psi}\|^2$ and $\|P_2 \ket{\psi}\|^2$, respectively, where $P_i$
is the projector onto the subspace $S_i$.  For a comprehensive
introduction to quantum computing, please refer to the textbooks
\cite{NC00, KSV02}.

\subsection{Definition of the problems}

The standard definition of Local Hamiltonian and Phase Estimation, two
problems extensively studied in quantum complexity theory and in
quantum algorithm engineering, are as follows.

\subsubsection*{Local Hamiltonian Minimal Eigenvalue (LH)}
\vspace{-.5em} \textbf{Input:} 1) Two numbers $a$ and $b$ such that
$a < b$ and the gap $b - a = \Omega(1/poly(n))$. 2) A Hamiltonian $H
= \sum_j H_j$ operating on $n$ qubits, where $j$ ranges over a set
of size polynomial in $n$, and each $H_j$ operates on a constant
number of qubits. It is promised that either $\lambda(H) < a$ or
$\lambda(H) > b$, where $\lambda(H)$ is the minimal eigenvalue of
$H$.

\vspace{.3em} \noindent \textbf{Output:} 0 if $\lambda(H) < a$, and
1 if $\lambda(H) > b$.

\subsubsection*{Phase Estimation (PE)}
\vspace{-.5em}\textbf{Input:} A unitary matrix $U$, given by
black-boxes of controlled-$U$, controlled-$U^{2^2}$, $\ldots$,
controlled-$U^{2^{t-1}}$ operations, and an eigenvector $\ket{u}$ of
$U$ with eigenvalue $e^{2\pi i \varphi}$ with the value of
$\varphi\in [0,1)$ unknown.

\vspace{.3em} \noindent\textbf{Output:} An $n$-bit estimation of
$\varphi$. \vspace{1em}

The paper will study several average and sampling variants of them.
For the sampling versions, we need a notion of the approximation of a
probability distribution. Suppose there is a probability distribution
$p$ on the sample space $X = \{x_1, ..., x_M\}$, on which a distance
function $d$ has been defined. Then a probability distribution $q$ on
$X$ is said to $(\epsilon, \delta)$-\emph{approximate} $p$ if there is
a decomposition $q_i = \sum_{j = 1}^{M} q_{ij}$ \st each $q_{ij} \geq
0$ and $\forall j$, $\sum_{i: d(x_i, x_j) \leq \epsilon} q_{ij} \geq
(1-\delta) p_j$. Intuitively, this means that if we draw a sample
according to $q$, then it will be $\epsilon$-close to $x_j$ with
probability at least $(1-\delta)p_j$. Sometimes we also say that a
sample drawn according to the probability distribution $q$ is \emph{an
estimation of $x_j$ up to $\epsilon$ with probability at least
$(1-\delta)p_j$}.

\vspace{.5em} \noindent \textbf{Comment.} \ Note that because of the
$\epsilon$-approximation (induced by the distance $d$), the standard
distance measures between two probability distributions, such as
total variance (or any $p$-norm generalization of it), Hellinger
distance, or Kullback-Leibler distance, do not qualify for our
purpose. For example, suppose $X = \{0.01, 0.02, ..., 1\}$, $p(x) =
0.1$ if $x \in \{0.1, 0.2, ..., 1\}$, and $q(x) = 0.1$ if $x \in
\{0.09, 0.19, ..., 0.99\}$. Then intuitively these two distributions
are very close to each other. But the total variance between them is
$\frac{1}{2}\sum_x |p(x) - q(x)| = 1$, the maximal possible
distance. (In this example, the Hellinger distance is 1, and the
Kullback-Leibler distance is infinity. All of these are not desired
because they treat different samples in the sample space as totally
different objects. But in many applications such as those in this
paper, the sample space has a natural distance measure on it.)
\vspace{.5em}

Now we are ready to define the sampling versions of LH and PE as
follows. Note that they are sampling problems rather than the
standard decision or searching problems: on every input, a sampling
problem is required to output some value with some probability.

\subsubsection*{Local Hamiltonian Eigenvalue Sampling (LHES)}
\vspace{-.5em} \textbf{Input:} 1) Hamiltonian $H = \sum_j H_j$
operating on $n$ qubits, where $j$ ranges over a set of size
polynomial in $n$, and each $H_j$ operates on a constant number of
qubits. Suppose the eigenvalues and the corresponding eigenvectors
of $H$ are $\{(\lambda_k, \ket{\eta_k})\}_k$ satisfying $|\lambda_k|
< poly(n)$ for each $k$. 2) An estimation precision $\epsilon =
\Omega(1/poly(n))$. 3) A sampling error probability $\delta =
\Omega(1/poly(n))$. 4) A classical $n$-bit string $b\in \B^n$.

\vspace{.3em} \noindent \textbf{Output:} An estimation of
$\lambda_k$ up to $\epsilon$ with probability at least
$(1-\delta)|\qip{b}{\eta_k}|^2$.

\subsubsection*{Phase Estimation Sampling (PES)}
\vspace{-.5em} \textbf{Input:} 1) $\langle U \rangle$, the
description of a $2^n\times 2^n$ unitary matrix $U$ (which is a
$poly(n)$-size quantum circuit). Suppose the eigenvalues of $U$ are
$\{\lambda_j = e^{2\pi i \varphi_j}\}_{j = 1, ..., 2^n}$ (where
$\varphi_j \in [0,1)$ for each $j$), with the corresponding
eigenvectors $\{\ket{\eta_j}\}_{j = 1, ..., 2^n}$. 2) An estimation
precision $\epsilon = \Omega(1/poly(n))$. 3) A sampling error
probability $\delta = \Omega(1/poly(n))$. 4) A classical $n$-bit
string $b\in \B^n$.

\vspace{.3em}\noindent \textbf{Output:} An estimation of $\varphi_j$
up to $\epsilon$ with probability at least
$(1-\delta)|\qip{b}{\eta_j}|^2$. \vspace{.5em}

One could also consider to estimate the average phase instead of
sampling the phase. One caveat here is that in the Phase Estimation
and the PES problems, it is implicitly assumed that the estimation is
in a circular sense, \ie if the $\varphi_u$ is very close to 1, then
an answer very close to 0 is considered correct.\footnote{This can be
seen from the analysis of the Phase Estimation algorithm; see, for
example, \cite{NC00}.} But when we consider the average phase, this
becomes a problem: if we have two unitary matrices, one with all
eigenvalues close to -1 and another with all eigenvalues close to 1,
then we may not be able to distinguish them since the average phase of
the second matrix could also be close to 1/2 due to the cancellation of
those small and large phases. For this reason, it is more natural to
consider the average eigenvalue of a unitary matrix. We define the
Local Unitary matrix Average Eigenvalue (LUAE) as follows.

\subsubsection*{Local Unitary matrix Average Eigenvalue
(LUAE)} \vspace{-.5em} \textbf{Input:} The same as that of PES.

\vspace{.3em}\noindent\textbf{Output:} An estimation of the average
eigenvalue $\overline{\lambda} = \sum_{j=1}^{2^n}
|\qip{b}{\eta_j}|^2 \lambda_j$ up to precision $\epsilon$ with
probability at least $1-\delta$. \vspace{.5em}

\vspace{1em} The main theorem of this paper can be written as
\begin{Thm}
LHES, PES and LUAE are all \bqp-complete.
\end{Thm}
\section{\bqp-hardness of the problems}\label{sec: hardness}
In this section we will show that the three problems LHES, PES and
LUAE are all \bqp-hard.

\subsection{LHES is \bqp-hard} \label{sec: LHES
reduction}
\begin{Thm}\label{thm: LHES reduction}
$\bqp \subseteq \bpp^{LHES}$
\end{Thm}
\begin{proof}
For any $L\in \bqp$, there is a uniform family of polynomial size
quantum circuits with $\epsilon$-bounded error (for a small constant
$\epsilon$) that decides if $x\in L$ or $x\not\in L$ as depicted in
Figure \ref{fig: U}. For $n$-bit inputs, denote by $U$ the
corresponding quantum circuit and suppose the size of $U$ is $M$,
which is bounded by a polynomial in $n$.  Further suppose that the
computation is described by $U\ket{x, \mathbf{0}} = \alpha_{x,0}
\ket{0}\ket{\psi_{x,0}} + \alpha_{x,1} \ket{1}\ket{\psi_{x,1}}$, where
$\mathbf{0}$ is the initial state of the ancillary qubits, and
$\ket{\psi_{x,0}}$ and $\ket{\psi_{x,1}}$ are pure states. After the
$U$ transform, the first qubit is measured and the algorithm outputs
the result based on the outcome of the measurement. As the requirement
of correctness, we have $|\alpha_{x,0}|^2 < \epsilon$ if $x\in L$, and
$|\alpha_{x,1}|^2 < \epsilon$ if $x\notin L$.

\begin{figure}[h]\label{fig: U}
\begin{center}
\epsfig{file=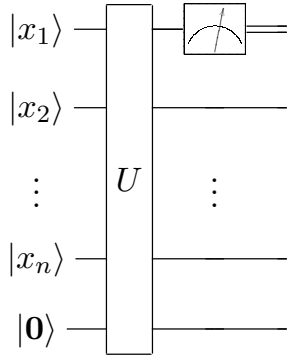, width=3.8cm} \caption{Circuit U} \label{fig: U}
\end{center}
\end{figure}

We now construct a local Hamiltonian $H$ encoding the circuit $U$ and
binary strings $b$ encoding the inputs $x$ such that eigenvalue
sampling applied to $H$ and $b$ yields significantly different
probability distributions for the two cases of $x\in L$ and $x\notin
L$. To this end, we construct the circuit $V$ in Figure \ref{fig:
VLH}, where we add a qubit $\ket{r}$ to store a copy of the output of
$U$.

\begin{figure}[h]
\begin{center}
\epsfig{file=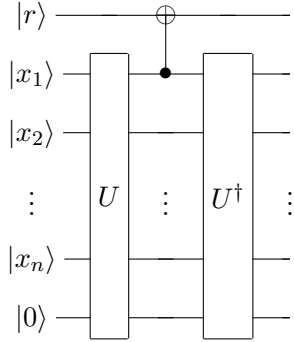, width=3.8cm} \caption{Circuit V} \label{fig:
VLH}
\end{center}
\end{figure}

Suppose the circuit is decomposed as $U = U_M \cdots U_1$, where each
$U_j$ an elementary gate. Then $U^\dag = U_1^\dag \cdots
U_M^\dag$. Let $V_j$ be the $j$-th gate in $V$, \ie $V_j = U_j$ for $j
= 1,\ldots,M$, $V_{M+1}$ be the CNOT gate, and $V_j = U_{2M+2-j}^\dag$
for $j = M+2,\ldots, 2M+1$. Let $N = 2M + 1$, the number of gates in
$V$. Attach a clock register $\ket{\cdot}_t$ to the system. Define the
operator
\begin{equation} \label{eq: def of F}
F = \sum_{j=1}^{N-1} V_j \otimes \ket{j}_t \bra{j-1}_t + V_N \otimes
\ket{0}_t\bra{N-1}_t.
\end{equation}
Note that $F$ is an $O(\log N)$-local operator; we will remark how to
slightly modify it to be a $4$-local operator at the end of the
proof. Define
\begin{equation}
\ket{\varphi_{x,j}} = F^j(\ket{0}\ket{x, \mathbf{0}}\ket{0}_t)
\end{equation}
for $j \geq 0$, where $F^j$ means that $F$ is applied $j$ times. Then
for $j = 0,...,2N-1$,
\begin{equation}
\ket{\varphi_{x,j}} =
 \left\{\!\!
\begin{array}{rll}
%
%
& \ket{0}\otimes U_j\cdots U_1\ket{x,\mathbf{0}}\otimes\ket{j}_t
& j=0,\ldots,M \vspace{.5em} \\
%
%
& \ket{0}\otimes U_{2M+2-j}^\dag \cdots U_M^\dag P_0
  U_M\cdots U_1\ket{x,\mathbf{0}}\otimes \ket{j}_t \\
+\!\!\!\!
& \ket{1}\otimes U_{2M+2-j}^\dag \cdots U_M^\dag P_1 U_M \cdots
  U_1\ket{x,\mathbf{0}}\otimes \ket{j}_t
& j = M+1,\ldots, 2M \vspace{.5em} \\
%
%
& \ket{0}\otimes U_{j-2M}^\dag \cdots
  U_M^\dag P_0 U_M \cdots U_1\ket{x,\mathbf{0}}\otimes\ket{j-N}_t \\
+\!\!\!\!
& \ket{1}\otimes U_{j-2M}^\dag \cdots U_M^\dag P_1
  U_M\cdots U_1\ket{x,\mathbf{0}}\otimes\ket{j-N}_t
& j = 2M+1,\ldots, 3M+1 \vspace{.5em} \\
%
%
& \ket{0}\otimes U_{2N-j} \cdots U_1\ket{x,\mathbf{0}}\otimes \ket{j-N}_t
& j = 3M+2,\ldots,4M+1
\end{array}
\right.
\end{equation}
where $P_0$ and $P_1$ are the projectors onto the subspaces of the
first output qubit of $U$ being 0 and 1, respectively. It is also
easy to see that $\ket{\varphi_{x,j}} = \ket{\varphi_{x,j \text{ mod
} 2N}}$ for $j\geq 2N$. 

Note that for different $j$ and $j'$ in $\{0, ..., 2N-1\}$,
$\ket{\varphi_{x,j}}$ and $\ket{\varphi_{x,j'}}$ are orthogonal if
$|j-j'| \neq N$ due to the clock register. Also note that
$U_M...U_1\ket{x,\mathbf{0}} = U\ket{x,\mathbf{0}} = \alpha_{x,0}
\ket{0}\ket{\psi_{x,0}} + \alpha_{x,1} \ket{1}\ket{\psi_{x,1}}$ by
definition. Therefore, we have
$\qip{\varphi_{x,j}}{\varphi_{x,N+j}} =
\bra{x,\mathbf{0}}U_1^\dag...U_M^\dag P_0
U_M...U_1\ket{x,\mathbf{0}} = |\alpha_{x,0}|^2$ for $j = 0,...,
N-1$. To summarize, we have
\begin{equation}\label{eq: inner product}
\qip{\varphi_{x,j}}{\varphi_{x,j'}} =
    \begin{cases}
    0 & |j - j'| \neq N \\
    |\alpha_{x,0}|^2 & |j-j'| = N
    \end{cases}
\end{equation}

Define the subspace $S_x = span\{\ket{\varphi_{x,j}}: j = 0, ...,
2N-1\}$. The key point here is that though $F$ has an exponentially
large dimension, $F|_{S_x}$ is of only polynomial dimension. Now if
$\alpha_{x,0} = 1$, then $S_x$ has dimension of $N$, and $F$ has a
period of $N$ on $S_x$. Actually, $F|_{S_x}$ is just a shift operator
on the basis $\{\ket{\varphi_{x,j}}\}_{j = 0,... N-1}$, \ie
$F\ket{\varphi_{x,j}} = \ket{\varphi_{x,j+1 \text{ mod } N}}$. It is
not hard to see that this operator has eigenvalues $\lambda_k =
\omega_N^k$ with corresponding eigenvectors $\ket{\xi_k} =
\frac{1}{\sqrt{N}} \sum_{j=0}^{N-1}
\omega_N^{-kj}\ket{\varphi_{x,j}}$, where $\omega_N = e^{2\pi i/N}$.
Also note the fact that
\begin{equation}\label{eq: prob for case 1}
|\qip{\varphi_{x,0}}{\xi_k}|^2 = 1/N
\end{equation}
for any $k = 0, ..., N-1$.

Now we consider the general case of $\alpha_{x,0} \neq 1$, in which
case $S_x$ has dimension $2N$ and $F$ has period of $2N$. Note that
if $x\notin L$, then $\ket{\varphi_{x,N+j}}$ is very close to
$\ket{\varphi_{x,j}}$; if $x\in L$, then $\ket{\varphi_{x,N+j}}$ is
almost orthogonal to $\ket{\varphi_{x,j}}$. See Figure \ref{fig: 2N
vectors} for an illustration.

\begin{figure}[h]
\begin{center}
\epsfig{file=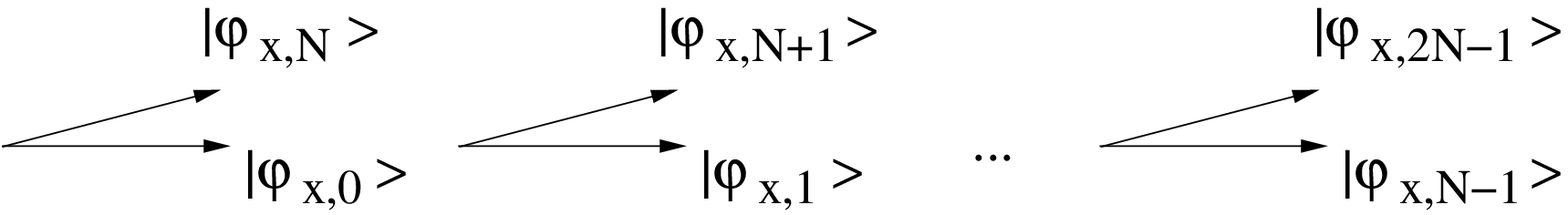, width=9cm}

(a) $x\notin L$ \vspace{2em}

\epsfig{file=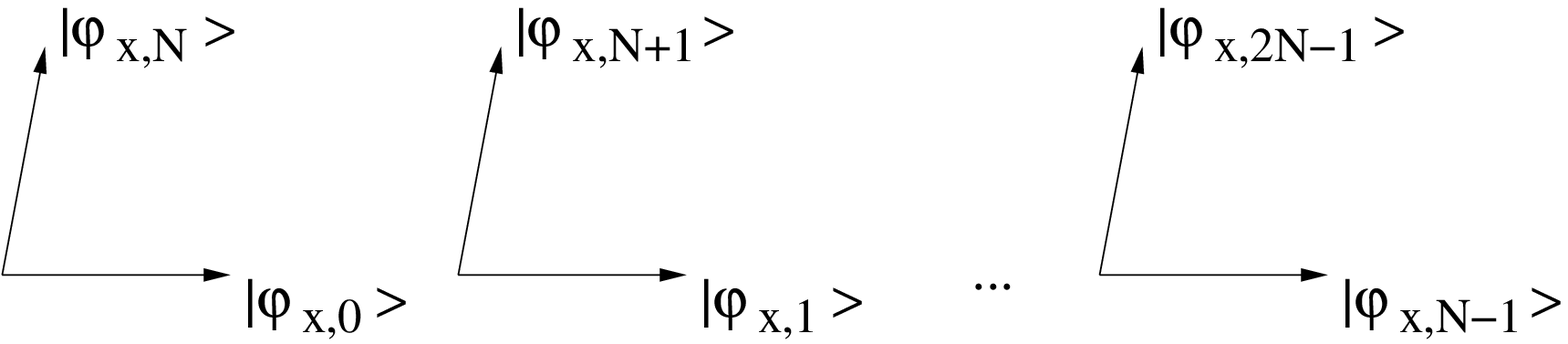, width=9cm}

(b) $x\in L$

\caption{The 2N vectors in the two cases} \label{fig: 2N vectors}
\end{center}
\end{figure}

To find the eigenvalues and eigenvectors of $F|_{S_x}$, define
\begin{equation}
\ket{\phi_{x,j}} = \frac{\ket{\varphi_{x,j}} +
\ket{\varphi_{x,N+j}}}{\|\ket{\varphi_{x,j}} +
\ket{\varphi_{x,N+j}}\|}, \quad \ket{\gamma_{x,j}} =
\frac{\ket{\varphi_{x,j}} - \ket{\varphi_{x,N+j}}}{
\|\ket{\varphi_{x,j}} - \ket{\varphi_{x,N+j}}\|}
\end{equation}
for $j = 0,...,N-1$. Then first, because
$\qip{\varphi_{x,j}}{\varphi_{x,N+j}} = |\alpha_{x,0}|^2$ is a real
number, we have $\qip{\phi_{x,j}}{\gamma_{x,j}} = 0$. Together with
Equality \eqref{eq: inner product}, we know that $\{\ket{\phi_{x,0}},
..., \ket{\phi_{x,N-1}}, \ket{\gamma_{x,0}}, ...,
\ket{\gamma_{x,N-1}}\}$ form an orthonormal basis of $S_x$.  Second,
since
\[
\ket{\varphi_{x,0}}    \xrightarrow{F}
\ket{\varphi_{x,1}}    \xrightarrow{F} \ldots \xrightarrow{F}
\ket{\varphi_{x,2N-1}} \xrightarrow{F} \ket{\varphi_{x,0}}\,,
\]
we observe that
\[
\ket{\phi_{x,0}} \xrightarrow{F} \ket{\phi_{x,1}} \xrightarrow{F}
\ldots \xrightarrow{F} \ket{\phi_{x,2N-1}} \xrightarrow{F}
\ket{\phi_{x,0}}
\]
and
\[
\ket{\gamma_{x,0}} \xrightarrow{F}
\ket{\gamma_{x,1}} \xrightarrow{F} ...  \xrightarrow{F}
\ket{\gamma_{x,2N-1}} \xrightarrow{F} -\ket{\gamma_{x,0}}\,.
\]
Therefore $F|_{S_x} = F_1 \oplus F_2$, where $F_1$ and $F_2$ act on
$S_{x,+} = span\{\ket{\phi_{x,0}}, ..., \ket{\phi_{x,N-1}}\}$ and
$S_{x,-} = span\{\ket{\gamma_{x,0}}, ..., \ket{\gamma_{x,N-1}}\}$,
respectively, with the matrix representations (in the basis
$\{\ket{\phi_{x,j}}\}$ and $\{\ket{\gamma_{x,j}}\}$, respectively) as
follows:
\begin{equation}
F_1 =
    \begin{pmatrix}
    0 & 0 & 0 & ... & 0 & 1 \\
    1 & 0 & 0 & ... & 0 & 0 \\
    0 & 1 & 0 & ... & 0 & 0 \\
    \vspace{1em}\vdots & \vdots & \vdots & & \vdots & \vdots \\
    0 & 0 & 0 & ... & 1 & 0
    \end{pmatrix},
\qquad F_2 =
    \begin{pmatrix}
    0 & 0 & 0 & ... & 0 & -1 \\
    1 & 0 & 0 & ... & 0 & 0 \\
    0 & 1 & 0 & ... & 0 & 0 \\
    \vspace{1em}\vdots & \vdots & \vdots & & \vdots & \vdots \\
    0 & 0 & 0 & ... & 1 & 0
    \end{pmatrix}.
\end{equation}
So $F_1$ is just a shift operator we mentioned before, which has
eigenvalues $\mu_k = \omega_N^k$ and eigenvectors $\ket{\eta_k} =
\frac{1}{\sqrt{N}} \sum_{j=0}^{N-1} w_N^{-kj} \ket{\phi_{x,j}}$. It is
also not hard to find that $F_2$ has the eigenvalues $\nu_k =
\omega_N^{k+1/2}$ with the eigenvectors $\ket{\varsigma_k} =
\frac{1}{\sqrt{N}} \sum_{j=0}^{N-1} w_N^{-(k+1/2)j}
\ket{\gamma_{x,j}}$. Let $P_+$ and $P_-$ be the projectors onto $S_+$
and $S_-$, respectively. Note that
\begin{equation}
\|P_- \ket{\varphi_{x,0}}\|^2 = \|\frac{\ket{\varphi_{x,0}} -
\ket{\varphi_{x,N}}}{2}\|^2 = \frac{2 -
\qip{\varphi_{x,0}}{\varphi_{x,N}} -
\qip{\varphi_{x,N}}{\varphi_{x,0}}}{4} =
\frac{1-|\alpha_{x,0}|^2}{2} = \frac{|\alpha_{x,1}|^2}{2}.
\end{equation}
Thus $\|P_+ \ket{\varphi_{x,0}}\|^2 = 1- \|P_-
\ket{\varphi_{x,0}}\|^2 = (1+|\alpha_{x,0}|^2)/2$. By this, we can
derive that
\begin{equation}\label{eq: prob in case 2}
|\qip{\varphi_{x,0}}{\eta_k}|^2 = \frac{1}{2N}(1 +
|\alpha_{x,0}|^2), \qquad |\qip{\varphi_{x,0}}{\varsigma_k}|^2 =
\frac{1}{2N}|\alpha_{x,1}|^2.
\end{equation}
for any $k = 0, ..., N-1$.

Now the local Hamiltonian is constructed as $H = F + F^\dag$. It is
easy to verify that $H$ is a local Hamiltonian. And further, suppose
the eigenvalues of $F$ are $\{\kappa_j\}$ with the eigenvectors
$\{\ket{\psi_j}\}$, then the eigenvalues of $H$ are just $\{\kappa_j
+ \kappa_j^*\}$ with the same corresponding eigenvectors. Now the
\bpp algorithm with the LHES oracle is as follows.

\vspace{1em} On input $x$,
\begin{enumerate}
\item \label{step: LHES oracle call}
Feed $(H, 1/4N, 1/100, 0x\mathbf{0}0)$ as input to the LHES oracle,
getting an output $a$

\item \label{step: range check}
    If $|a| > 1$,

    \quad go back to Step \ref{step: LHES oracle call};

\item If $\min_{k=0,...,N-1}{|a-2\cos\frac{2\pi (k+1/2)}{N}}| <
    \min_{k=0,...,N-1}{|a-2\cos\frac{2\pi k}{N}}| $

    \quad output ``$x\in L$"

    else

    \quad output ``$x\notin L$"

\end{enumerate}

The algorithm is correct with probability at least
$\frac{1}{2}(1-\delta)(1-\epsilon)$ for $x\in L$, and correct with
probability at least $(1-\delta)(1-\frac{\epsilon}{2})$ for $x\notin
L$ . Actually, in Step \ref{step: LHES oracle call}, the oracle will
output an estimation of $2\cos \frac{2\pi k}{N}$ (up to an additive
$\frac{1}{4N}$) with probability at least
$\frac{1}{2N}(1-\delta)(1+|\alpha_{x,0}|^2)$, and output an
estimation of $2\cos \frac{2\pi (k+1/2)}{N}$ (up to an additive
$\frac{1}{4N}$) with probability at least
$\frac{1}{2N}(1-\delta)|\alpha_{x,1}|^2$, for any $k = 0, ..., N-1$.
Note that now the case of $\alpha_{x,0} = 1$ can be consistently
viewed as a special case of $\alpha_{x,0} \neq 1$ by comparing
Equality \eqref{eq: prob for case 1} and \eqref{eq: prob in case 2}.

If $k\in [N/6, N/3]\cup [2N/3, 5N/6]$, which happens with
probability 1/3 for a uniformly random $k\in \{0, ..., N-1\}$, then
the algorithm will proceed to the outer ``else" branch in Step
\ref{step: range check}. Note that for $\theta, \theta' \in [\pi/3,
2\pi/3] \cup [4\pi/3, 5\pi/3]$ and $|\theta - \theta'| =
\frac{1}{2N}$, we have $|2\cos\theta - 2\cos\theta'| \geq
\frac{\sqrt{3}}{2N}
> \frac{1}{2N}$. So if $x\in L$, then with at least probability
$\frac{1}{2}(1-\delta)|\alpha_{x,1}|^2 \geq
\frac{1}{2}(1-\delta)(1-\epsilon) $, we can get a value $``\leq
\frac{1}{4N}$-close" to $2\cos\frac{2\pi (k+1/2)}{N}$ for some $k$,
but ``$> \frac{1}{4N}$-far" from any $2\cos\frac{2\pi k}{N}$, so the
last step will catch this and output the correct answer.

The case of $x\notin L$ can be similarly analyzed. Also, it is easy to
see that the expected running time of the algorithm is polynomial in
the input size, which completes the proof for the $O(\log n)$-local
LHES reduction.

Finally, to obtain a $4$-local LHES, we replace $\ket{i}$ by
$\ket{e_i}=\ket{0...010...0}$ for $i=0\ldots,N-1$, where the only $1$
appears at coordinate $i$. Modify the operator $F$ to be
\begin{equation}
F = \sum_{j = 1}^ {N-1} V_j \otimes \ket{e_j}\bra{e_{j-1}} + V_N
\otimes \ket{e_0}\bra{e_{N-1}}\,.
\end{equation}
Note that $\ket{e_j}\bra{e_{j-1}}$ and $\ket{e_0}\bra{e_{N-1}}$ are
2-local. The remaining proof passes through.
\end{proof}

\subsection{PES and LUAE are \bqp-hard}
\begin{Thm}\label{thm: PES reduction}
$\bqp \subseteq \bpp^{PES}$, $\bqp \subseteq \bpp^{LUAE}$.
\end{Thm}

\begin{proof}
For any $L\in \bqp$, there is a uniform family of polynomial size
quantum circuits with $\epsilon$-bounded error (for a small constant
$\epsilon$) that decides if $x\in L$ or $x\not\in L$ as in Figure~
\ref{fig: U}. For any input $x\in \B^n$, suppose the corresponding
circuit $U$ operates on $N = poly(n)$ qubits, taking $x$ and some
ancillary bits $\mathbf{0} = 0^{N-n}$ as the input.  After the circuit
computation, measuring the first qubit of the output
$U\ket{x}\ket{\mathbf{0}}$ give the correct result with probability
great than $1-\epsilon$. Suppose $U\ket{x}\ket{\mathbf{0}} =
\alpha_{x,0} \ket{0}\ket{\psi_{x,0}} + \alpha_{x,1}
\ket{1}\ket{\psi_{x,1}}$, then $|\alpha_{x,0}|^2 < \epsilon$ if $x\in
L$, and $|\alpha_{x,1}|^2 < \epsilon$ if $x\notin L$.

Now we construct another circuit $V$ as in Figure \ref{fig: VPE},
where $Z$ is the Pauli-Z matrix
$\begin{bmatrix}
1 & 0 \\ 0 & -1
\end{bmatrix}$.

\begin{figure}[h]
\begin{center}
\epsfig{file=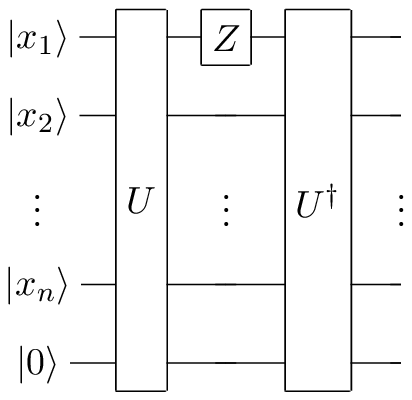, width=3.8cm} \caption{Circuit V} \label{fig:
VPE}
\end{center}
\end{figure}

It is easy to see that
\begin{align}
V\ket{x, \mathbf{0}} & = U^\dag (\alpha_{x,0}\ket{0}\ket{\psi_{x,0}}
- \alpha_{x,1} \ket{1}\ket{\psi_{x,1}}) \\
\label{eq: V transform} & = \begin{cases}
    \quad   \ket{x,\mathbf{0}} - 2\alpha_{x,1}U^\dag\ket{1}\ket{\psi_{x,1}} \\
    \, - \, \ket{x,\mathbf{0}} + 2\alpha_{x,0}U^\dag\ket{0}\ket{\psi_{x,0}}
    \end{cases}
\end{align}
Therefore, $\ket{x,\mathbf{0}}$ is almost an eigenvector of the
eigenvalue $-1$ or $1$, depending on whether $x\in L$ or not. So we
can use the PES or LUAE oracle to distinguish between these two
cases.

\vspace{.5em} \noindent 1. $\bqp \subseteq \bpp^{PES}$: We run the
PES oracle on input $(\langle V \rangle, 1/8, 1/100, x0^{N-n})$, and
the oracle outputs a (sampled) value $\theta$. Accept if $1/4 \leq
\theta \leq 3/4$, and reject otherwise.

To see why this is correct, write $V$ as $V =
\sum_j\lambda_j\ket{\eta_j}\bra{\eta_j}$, where $\lambda_j = e^{2\pi
i \varphi_j}$ are eigenvalues and $\ket{\eta_j}$ are the
corresponding eigenvectors. Consider $x\notin L$ first, and the
other case $x\in L$ is symmetric. Denote $x0^{N-n}$ by $b$ and
decompose it as $\ket{b} = \sum_j \beta_j \ket{\eta_j}$. Then
$V\ket{b} - \ket{b} = \sum_j (\lambda_j - 1) \beta_j\ket{\eta_j}$.
By \eqref{eq: V transform}, this implies
\begin{equation} \label{eq: error eq}
\sum_j (\lambda_j - 1) \beta_j\ket{\eta_j} = -
2\alpha_{x,1}U^\dag\ket{1}\ket{\psi_{x,1}}.
\end{equation}
Since the PES oracle outputs an estimation of $\varphi_j$ up to 1/8
with probability at least $(1-\delta)|\qip{b}{\eta_j}|^2$ with
$\delta = 1/100$, the success probability of our algorithm is
\begin{align}
\pr[\theta < 1/4 \text{ or } \theta > 3/4] & \geq  \sum_{j:
\varphi_j < 1/8 \text{ or } \varphi_j > 7/8} (1-\delta) |\qip{b}{\eta_j}|^2 \\
& = (1-\delta) \sum_{j: |\lambda_j - 1| < 2\sin(\pi/8)} |\beta_j|^2 \\
& = (1-\delta) (1-\sum_{j: |\lambda_j - 1| \geq 2\sin(\pi/8)} |\beta_j|^2) \\
& \geq (1-\delta) (1-\sum_{j: |\lambda_j - 1| \geq 2\sin(\pi/8)} \frac{|\lambda_j - 1|^2}{4\sin^2(\pi/8)}|\beta_j|^2) \\
& \geq (1-\delta) (1-\sum_{j} \frac{|\lambda_j - 1|^2}{4\sin^2(\pi/8)}|\beta_j|^2) \\
& = (1-\delta) (1-\frac{1}{4\sin^2(\pi/8)}\| \sum_j (\lambda_j - 1) \beta_j\ket{\eta_j} \|^2) \\
& = \label {eq: a} (1-\delta) (1-\frac{1}{4\sin^2(\pi/8)}\| 2\alpha_{x,1}U^\dag\ket{1}\ket{\psi_{x,1}}\|^2) \\
& = (1-\delta) (1-\frac{|\alpha_{x,1}|^2 }{\sin^2(\pi/8)} ) \\
& = (1-\delta) (1-\frac{\epsilon }{\sin^2(\pi/8)} )
\end{align}
where \eqref{eq: a} is because of equality \eqref{eq: error eq}.
Therefore the error probability is less than $O(\epsilon)$. The same
arguments can show the correctness for the case of $x\in L$, which
completes our proof for the first part.

\vspace{.5em} \noindent 2. $\bqp \subseteq \bpp^{LUAE}$: For the
circuit $V$ in Figure \ref{fig: VPE}, it is easy to show that
$\overline{\lambda}$ is close to $-1$ (and $1$) if $x\in L$ (and
$x\notin L$). First observe that for any $b$, we have
$\overline{\lambda} = \bra{b}U\ket{b}$, which can be shown by
writing both $U$ and $b$ in terms of $\ket{\eta_j}$'s. Now for $b =
x\mathbf{0}$, if $x\notin L$, then $|\overline{\lambda} -1 | =
|\bra{b} U \ket{b} -1 | = |\bra{b} U \ket{b} - \qip{b}{b} | \leq
|U\ket{b} - \ket{b}| = 2|\alpha_{x,1}| \leq 2\sqrt{\epsilon}$.
Similarly if $x\in L$, then $|\overline{\lambda} + 1 | \leq
2\sqrt{\epsilon}$. Thus a good estimation to $\overline{\lambda}$
suffices to distinguish between $x\in L$ and $x\notin L$.
\end{proof}

\section{The problems are in \bqp}\label{sec: in BQP}
In this section, we will prove that LHES, PES and LUAE are all in
\bqp. We will first review the standard algorithm for Phase
Estimation, then observe that the same algorithm actually gives the
desired PES solution. We then use it to show an algorithm for
LHES. Finally, we prove that LUAE is in \bqp.

\subsection{Phase Estimation and an efficient quantum algorithm for PES}
Phase Estimation can be solved by a quantum algorithm as follows
(see, for example, \cite{NC00}). The working space has two
registers. The first register consists of $t = n +
\ceil{\log(2+1/2\delta)}$ qubits and is prepared in $\ket{0\ldots
0}$. The second register contains the eigenvector $\ket{u}$.
Measuring $\tilde{\varphi}$ in the first register after carrying out
the transformations described below gives the desired $n$-bit
estimation of $\varphi$ with probability of at least $1-\delta$.
\begin{align}
& \ket{0}^{\otimes t}\ket{u} \\
\rightarrow & \frac{1}{\sqrt{2^t}}\sum_{j=0}^{2^t-1}\ket{j}\ket{u}
& // \text{ apply the Fourier transform}  \\
\rightarrow & \frac{1}{\sqrt{2^t}}\sum_{j=0}^{2^t-1}\ket{j}U^j
\ket{u} & // \text{ apply the controlled powers of $U$} \\
= & \frac{1}{\sqrt{2^t}}\sum_{j=0}^{2^t-1}e^{2\pi i j
\varphi}\ket{j} \ket{u} \\
\rightarrow & \ket{\tilde{\varphi}}\ket{u} & // \text{ apply the
inverse Fourier transform}
\end{align}

\vspace{1em} The following observation says that the same algorithm
actually works for PES.

\vspace{1em}\noindent\textbf{Fact.} If we feed $\ket{0}\ket{b}$
instead of $\ket{0}\ket{u}$ as input to the above algorithm for the
Phase Estimation problem and let $t = \ceil{\log \frac{1}{\epsilon}} +
\ceil{\log(2+\frac{1}{2\delta})}$, then the measurement of the first
register gives the desired sampling output. This implies that PES can
be solved by a \bqp machine.

\vspace{1em} This actually holds not only for $\ket{b}$ but also for a
general state $\ket{\eta}$. To see why this is true, write
$\ket{\eta}$ as $\sum_{j=1}^n \alpha_{j}\ket{\eta_j}$, then by the
linearity of the operations, the final state is
$\alpha_{j}\ket{\tilde{\varphi_j}}\ket{\eta_j}$. For more details, we
refer the readers to \cite{NC00} (Chapter 5). Note that to implement
the controlled-$U^{2^j}$ operations for $j = 0, ..., 2^t-1$ in the
above algorithm, we need to run $U$ for $2^t$ times, which can be done
efficiently since $t = \ceil{\log \frac{1}{\epsilon}} +
\ceil{\log(2+\frac{1}{2\delta})}$ and $\epsilon = \Omega(1/poly(n))$,
$\delta = \Omega(1/poly(n))$.

\subsection{LHES in \bqp} \label{sec: LHES in BQP}
\begin{Thm}\label{thm: LHES in BQP}
LHES can be implemented by a uniform family of quantum circuits of
polynomial size.
\end{Thm}
\begin{proof}
By a simple scaling ($H' = H/\Lambda = \sum_j H_j/\Lambda$ where
$\Lambda = \max_k |\lambda_k| = poly(n)$), we can assume that all
the eigenvalues $\lambda_k$ of $H$ satisfy $ |\lambda_k| < 1/4$. The
basic idea to design the \bqp algorithm is to use Phase Estimation
Sampling on $e^{2\pi i H}$. Note that $e^{2\pi i H}$ is unitary, and
if the eigenvalues and eigenvectors of $H$ are $\{\lambda_k,
\ket{\eta_k}\}$, then those of $e^{2\pi i H}$ are just $\{e^{2\pi i
\lambda_k}, \ket{\eta_k}\}$. Therefore, it seems that it is enough
to run the PES algorithm on $(e^{2\pi i H},\epsilon, \delta, b)$,
and if we get some $\lambda > 1/2$, then output $\lambda - 1$.
However, note that $H$ is of exponential dimension, so $e^{2\pi i
H}$ is not ready to compute in the straightforward way. Fortunately,
this issue is well studied in the quantum simulation algorithms, and
the standard approach is the following asymptotic approximation by
the Trotter formula \cite{Tr59, Ch68, NC00} or its variants. Here
using the simulation technique, we obtain
\begin{equation}
\left(e^{2\pi i \sum_j H_j/m}\right)^m = \left(\prod_j e^{2\pi i
H_j/m}\right)^m + O(1/m)
\end{equation}
Now we run PES on $(b, \epsilon, \delta/2, e^{2\pi i H})$. Whenever
we need to call $e^{2\pi i H}$, we use $\prod_j e^{2\pi i H_j/m}$
for $m$ times instead. Note that such substitution yields $O(1/m)$
deviation, so $t = \log \frac{2}{\epsilon\delta} + O(1)$ calls yield
$O(\frac{1}{m\epsilon\delta}) \leq \frac{c}{m\epsilon\delta}$
deviation for some constant $c$. Let $m =
\frac{2c}{\epsilon\delta^2}$, thus the final error probability is
less than $\frac{\delta}{2} + \frac{c}{m\epsilon\delta} = \delta$,
achieving the desired estimation and sampling precisions.
\end{proof}

\vspace{1em} \noindent \textbf{Comment.} From the proof we can see
that as long as $e^{2\pi i H}$ can be simulated efficiently from the
description of $H$, we can sample the eigenvalues of $H$ as
desired. Since sparse Hamiltonians, which contains local Hamiltonians
as special cases, can be simulated efficiently \cite{AT03, BACS05}, we
know that if we modify the definition of LHES by allowing $H$ to be
sparse, then it is also \bqp-complete.

\subsection{Local unitary matrices average eigenvalue
estimation}\label{sec: LUAE}
\begin{Thm}\label{thm: LUAE in BQP}
LUAE is in \bqp.
\end{Thm}
We can use the \bqp algorithm for PES to get $O(1/(\epsilon \delta)) $
(independent) samples $\varphi_{(1)}, ..., \varphi_{(m)}$ and use the
sample mean $\widehat{\lambda} = \frac{1}{m}\sum_{j=1}^m e^{2\pi i
\varphi_{(j)}}$ as an desired estimation of $\overline{\lambda}$,
which can be proved by studying the Phase Estimation algorithm in more
details \footnote{Directly applying the definition of the PES problem
as a black-box is not enough: the mean of the output of the PES
algorithm may be far away from $\overline{\lambda}$ because of the
exponentially many eigenvalues and for each eigenvalue the output
probability is $\epsilon$ away from the correct one.}. But here we
shall give a different algorithm whose analysis is much
simpler. Recall that $\overline{\lambda} = \bra{b}U\ket{b}$. So after
we apply the circuit $U$ to $\ket{b}$, the problem becomes to estimate
the inner product of two quantum states $\ket{b}$ and $U\ket{b}$,
which can be done by the standard SWAP test. For example, Yao
\cite{Yao03} gave an estimate of $|\qip{u}{v}|$ up to $\epsilon$ with
error probability $\delta$ by applying the SWAP test in \cite{BCWdW01}
on $O(\frac{1}{\epsilon^4}\log \frac{1}{\delta})$ pairs of $(\ket{u},
\ket{v})$. Here for the special two states $\ket{b}$ and $U\ket{b}$,
it turns out that we can slightly decrease the cost to
$O(\frac{1}{\epsilon^2}\log \frac{1}{\delta})$, by using the following
lemma proved in \cite{WY06}:
\begin{Lem} \label{lem: sampling
lemma} Let $U$ be a quantum circuit of length $O(poly(n))$ acting on
n qubits, and let $\ket{\psi}$ be a pure state of the n qubits which
can be prepared in time $O(poly(n))$. It is then possible to sample
from random variables $X, Y \in \{-1, +1\}$ for which
\begin{equation}
\av[X + iY ] = \bra{\psi}U\ket{\psi}
\end{equation}
in $O(poly(n))$ time.
\end{Lem}
\begin{proof} (of Theorem \ref{thm: LUAE in BQP}) Run the sampling
algorithm in the lemma to get $m = O(\frac{1}{\epsilon^2}\log
\frac{1}{\delta})$ samples $(X_1, Y_1), ..., (X_m, Y_m)$. Let
$\overline{X} = \frac{1}{m}\sum_{j=1}^m X_j$ and $\overline{Y} =
\frac{1}{m}\sum_{j=1}^m Y_j$. Use the sample mean $\widehat{\lambda}
= \overline{X} + i \overline{Y} $ as an desired estimation of
$\overline{\lambda}$. Noting $\var[X] \leq 4$ and $\var[Y] \leq 4$
(by the definition of variance), we have by Chernoff's bound that
\begin{align}
\pr[|\widehat{\lambda} - \overline{\lambda} | \geq \epsilon] \leq
 \pr[|\widehat{X} - Re(\overline{\lambda})| \geq \epsilon/2] +
 \pr[|\widehat{Y} - Im(\overline{\lambda})| \geq \epsilon/2] \leq \delta.
\end{align}
as desired.
\end{proof}

\section{Discussions and open problems}\label{sec: conclusion}
One might wonder why not to consider the average eigenvalue of a
local Hamiltonian. Actually, since $\bra{b}H\ket{b} = \bra{b}\sum_j
H_j\ket{b} = \sum_j \bra{b}H_j\ket{b}$ and each $\bra{b}H_j\ket{b}$
can be easily computed even deterministically, we can obtain the
exact average eigenvalue of a local Hamiltonian deterministically in
polynomial time.

However, as shown very recently by Janzing and Wocjan, if we
generalize the problem to estimating the average eigenvalue of $H^m$
where $m = poly(n)$ is part of the input, then the problem of
estimating $\bra{b}H^m\ket{b}$ is $\bqp$-complete \cite{JW06}.

We can also consider the complexity of unguided version of the
problems studied so far. That is, $b$ is not part of the input, and
we want to sample the eigenvalue/phase with equal probability or
estimate the average eigenvalue under the uniform distribution over
all the $2^n$ eigenvalues. We use $LHES_u, PES_u, LUAE_u$ to denote
the corresponding problems. We basically do not know anything about
these problems yet, except some trivial facts like $LUAE_u \in \bqp$
due to the following simple observation: $\av_b[\bra{b}U\ket{b}] =
\frac{1}{2^n}\sum_b \bra{b}U\ket{b} =
\frac{1}{2^n}\sum_{b,j}|\qip{b}{\eta_j}|^2 \lambda_j =
\frac{1}{2^n}\sum_j\lambda_j = \bar{\lambda}$. (We can just
uniformly choose $b\in \B^n$ and run the algorithm in Lemma
\ref{lem: sampling lemma}.)

\vspace{2em} \noindent \textbf{Acknowledgments}

We would like to thank John Preskill and Dominik Janzing for useful
discussions and Andrew Yao for comments on the manuscript. Thanks also
to Yaoyun Shi for reading a preliminary version and asking about the
complexity of estimating the average phase of a unitary matrix, which
led us to consider the problem LUAE.

\end{document}